\documentclass[submission, Phys]{SciPost}

\usepackage[bitstream-charter]{mathdesign}
\DeclareSymbolFont{usualmathcal}{OMS}{cmsy}{m}{n}
\DeclareSymbolFontAlphabet{\mathcal}{usualmathcal}

\usepackage{graphicx}% Include figure files
\usepackage{dcolumn}% Align table columns on decimal point
\usepackage{bm}% bold math

\usepackage{braket}
\usepackage{siunitx}
\usepackage{xcolor}

\hypersetup{
    colorlinks,
    linkcolor={red!50!black},
    citecolor={blue!50!black},
    urlcolor={blue!80!black}
}

\newcommand{\El}{\mathcal{E}}
\newcommand{\locOp}{\mathcal{O}}
\newcommand{\ii}{\mathrm{i}}    %complex i
\newcommand{\pd}{\partial{}}    %partial d
\newcommand{\veca}{\boldsymbol{\alpha}}         %bold \alpha
\newcommand{\vecat}{\boldsymbol{\alpha}(t)}     %bold \alpha(t)
\newcommand{\vecr}{\boldsymbol{x}}         %bold x

\newif\ifcomments\commentstrue
\newcommand{\hide}[1]{}

\newcommand{\figref}[1]{Fig.~\ref{#1}}

\newcommand{\JKU}{Institute for Theoretical Physics, Johannes Kepler
University Linz, Altenberger Straße 69, 4040 Linz, Austria}
\newcommand{\UPC}{Departament de F\'{\i}sica i Enginyeria Nuclear, Campus
Nord B4-B5, Universitat Polit\`ecnica de Catalunya, E-08034 Barcelona, Spain}

\begin{document}

% TODO: write your article's title here.
% The article title is centered, Large boldface, and should fit in two lines
\begin{center}{\Large \textbf{
Time-dependent variational Monte Carlo study of the dynamic response
of bosons in an optical lattice\\
}}\end{center}

% TODO: write the author list here. Use first name (+ other initials) + surname format.
% Separate subsequent authors by a comma, omit comma and use "and" for the last author.
% Mark the corresponding author with a superscript star.
\begin{center}
Mathias Gartner\textsuperscript{1$\star$},
Ferran Mazzanti\textsuperscript{2} and
Robert E. Zillich\textsuperscript{1}
\end{center}

% TODO: write all affiliations here.
% Format: institute, city, country
\begin{center}
{\bf 1} \JKU{}
\\
{\bf 2} \UPC{}
\\
% TODO: provide email address of corresponding author
${}^\star${\small \sf mathias.gartner@jku.at}
\end{center}

\begin{center}
\today
\end{center}

% For convenience during refereeing (optional),
% you can turn on line numbers by uncommenting the next line
%\linenumbers
% You should run LaTeX twice in order for the line numbers to appear.

\section*{Abstract}
\textbf{
We study the dynamics of a one-dimensional Bose gas at unit filling in both
shallow and deep optical lattices and obtain the dynamic structure factor
$\bm{S(k,\omega)}$ by monitoring the linear response to a weak probe pulse.
We introduce a new procedure, based on the
time-dependent variational Monte Carlo method (tVMC), which allows to evolve
the system in real time, using as a variational model a Jastrow-Feenberg
wave function that includes pair correlations.
Comparison with exact diagonalization results of $\bm{S(k,\omega)}$ obtained on a
lattice in the Bose-Hubbard limit shows good agreement of the dispersion
relation for sufficiently deep optical lattices, while for shallow lattices
we observe the influence of higher Bloch bands. 
We also investigate non-linear response to strong pulses. From the power spectrum of the density
fluctuations we obtain the excitation spectrum,
albeit broadened, by higher harmonic generation after a strong pulse with a
single low wave number.
As a remarkable feature of our simulations we furthermore demonstrate that the full
excitation spectrum can be retrieved from the power spectrum of the density fluctuations due to
the stochastic noise inherent in
any Monte Carlo method, without applying an actual perturbation.
}

% TODO: include a table of contents (optional)
% Guideline: if your paper is longer that 6 pages, include a TOC
% To remove the TOC, simply cut the following block
\vspace{10pt}
\noindent\rule{\textwidth}{1pt}
\tableofcontents\thispagestyle{fancy}
\noindent\rule{\textwidth}{1pt}
\vspace{10pt}

\section{Introduction} %\the\columnwidth{}
The dynamic structure factor $S(k,\omega)$ is a fundamental quantity as it
contains the maximal information about the dynamics of many-body quantum systems
that one can obtain by inelastic scattering~\cite{Newton2002}, such as the
excitation energies $\omega(k)$ and the lifetime of collective excitations. 
In quantum gases $S(k,\omega)$ can be measured by Bragg
spectroscopy \cite{stenger1999prl}, with relative momentum and energy resolution
similar to inelastic neutron scattering in condensed matter~\cite{sobirey2021s}.
The calculation of $S(k,\omega)$ is a demanding task beyond very simple
Hamiltonians or approximations, such as the Bijl-Feynman
model~\cite{bijl1940p,feynman1956pr}, or the Bogoliubov-de Gennes technique in
the mean field limit~\cite{pitaevskii2016, dalfovo1999rmp}.
Advanced variational methods based on action minimization, 
such as the correlated basis function approach~\cite{campbell2015prb} 
or the multi-configuration time-dependent Hartree algorithm~\cite{lode2020rmp},
can achieve much more accurate results. 
All these methods are reliable in many cases but are not expected to work well
in all situations.
Monte Carlo methods, on the other hand, are known to be able to produce
statistically exact predictions for bosons, although this only applies to the
ground state at zero temperature~\cite{kosztin1996ajop}, or to static ensemble
averages at finite temperature~\cite{ceperley1995rmp}. 
Access to the excitation spectrum is restricted to the evaluation of the dynamic
response in imaginary time and its reconstruction in frequency space by
inverting the Laplace transform. This is a rather difficult procedure as Laplace
inversion is a well known ill-posed mathematical problem, worsened in practice
by the fact that the stochastic noise of the simulation is exponentially
amplified in the result.
The way to tackle these problems is to build many reconstructions of the
response and to use stochastic methods based on simulated
annealing~\cite{rota2015jcp} or genetic
algorithms~\cite{vitali2010prb,bertaina2016prl} to produce 
an approximate dynamic structure factor.
While this method can yield good results, it is computationally very expensive
and usually gets only the broad features, not resolving well the fine details of
the response.
Other methods available for dynamic simulations are either restricted to lattice
systems, like time-evolving block decimation~\cite{vidal2004prl}, nonequilibrium
dynamical mean-field theory ~\cite{aoki2014rmp}, and the time-dependent density
matrix renormalization group
method~\cite{white1992prl,feiguin2005prb,schollwock2005rmp}, or they work best
in one dimension, like methods based on continuous matrix product
states~\cite{verstraete2010prl}.
Consequently, accurate methods that allow for time dependent simulations of
strongly correlated many-body systems which can describe the linear, but also
nonlinear response to perturbations, are in demand.

The time-dependent variational Monte Carlo (tVMC) method
\cite{carleo2012sr,carleo2014pra,carleo2017prx,dawid2022} is particularly suitable
for the study of quantum many-body dynamics, allowing for perturbations of any
strength. It can be applied to analyze many different situations, such as
ramping up the lattice depth~\cite{gardas2017prb} or interaction
quenches~\cite{carleo2014pra}, as well as many-body dynamics far from
equilibrium~\cite{eisert2015np}. 
It has also been extended to wave functions based
on artificial neural networks~\cite{carleo2017s,schmitt2020prl}.
In this work we use tVMC to analyze the dynamic response of a Bose gas to a probe
pulse in an optical lattice in one dimension, where we use a continuous space representation
rather than the Bose-Hubbard limit. We present a new way to calculate the dynamic
structure factor $S(k,\omega)$ of strongly interacting bosons in continuous space,
based on tVMC simulations of the time evolution after weak pulses. For strong pulses, we
enter the nonlinear regime. A strong perturbation with only a single wave number creates
excitations with multiplies of this wave number due to higher harmonic generation.
We exploit this to obtain a broadened approximation of the full excitation spectrum
by analysing the power spectrum of the density fluctuations after such strong pulse
with a single low wave number. Finally, we introduce a third way to calculate
the excitations with tVMC: surprisingly, we can obtain the excitation spectrum from
the power spectrum of the density fluctuations with no perturbation at all,
i.e. from the tVMC time evolution of the variational ground state, thanks to
the stochastic noise inherent in Monte Carlo simulations.

\section{Method}

We use tVMC
to study the response of the Bose gas, initially in the ground state at time $t=0$, to
an external perturbation $\delta V_p(x,t)$, and monitor the time
evolution of the density fluctuations $\delta \rho(x, t)=\rho(x,t)-\rho(x,0)$.
In the linear response regime, the ratio of their Fourier transforms
is the density response function, with its imaginary part being the dynamic
structure factor~\cite{pines1966}.
We perform a series of simulations for a system of $N$ identical bosons of mass~$m$, moving in a one dimensional optical lattice $V(x)$ and interacting via a contact
potential.
The Hamiltonian reads
\begin{equation}
    H = \sum_{i=1}^N \left( -\frac{\hbar^2}{2m}\frac{\pd^2}{\pd x_i^2} + V(x_i) + 
    \delta V_p(x_i,t) \right) + g \sum_{i<j}^N \delta(x_i - x_j) \,,
    \label{eq:hamiltonian}
\end{equation}
with the coupling constant $g$ parametrizing the strength of the two-body
interaction. 
As usual for cold atomic systems, where the optical lattice potential is generated by counter propagating laser beams with wave number $k_L$, we will use the form $V(x) = V_0 \sin^2(k_L x)$ for the potential~\cite{bloch2008rmp}, which corresponds to a lattice constant of $\pi/k_L$.
Throughout this work we will report $V_0$ and~$g$
in units of the recoil energy ${E_r = \hbar^2 k_L^2/2m}$
and ${E_r/k_L}$, 
respectively,
and we use $x_0=\pi/k_L$ and 
$t_0=\hbar/E_r$ as length and time unit.

%###############################################################################
In the deep lattice limit where the amplitude $V_0$ is large, $H$ can be
approximated by the lattice Hamiltonian of the single-band Bose-Hubbard model
(BHM)~\cite{hubbard1963potrsolsamaps, fisher1989prb}
\begin{equation}
    H_{\text{BHM}}=-J \sum_{i<j}^N b_i^\dagger b_j + U/2 \sum_i^N n_i (n_i - 1) \,,
\end{equation}
where $b_i^\dagger$, $b_i$ and $n_i$ are the creation, annihilation and number
operator for bosons at lattice site $i$.
For given $V_0$ and $g$ in Eq.~\eqref{eq:hamiltonian}, the on-site
interaction~$U$ and the hopping parameter~$J$ of the BHM can be evaluated
numerically performing band-structure calculations~\cite{jaksch1998prl}.
Within our continuous space tVMC simulations, we can access both the BHM regime
and the region of shallow optical lattices, where the single-band BHM is
no longer valid.

%###############################################################################
\subsection{Model wavefunction}

The tVMC method relies on a model wave function with variational parameters that
are propagated in time.
For modeling the time-dependent wavefunction $\Phi(\vecr, t)$ of the many-body
system, with ${\vecr=(x_1,\dots,x_N)}$, we use a Jastrow-Feenberg
ansatz~\cite{feenberg1969} with one- and two-particle correlation functions
\begin{equation}
    \Phi(\vecr, t) = e^{\sum_i^N u_1(x_i, t)}\, e^{\sum_{i, j}^N u_2(x_i - x_j, t)} \,.
    \label{eq:phi0new}
\end{equation}
In tVMC simulations, we parametrize the wavefunction by a set of time-dependent complex variational parameters ${\veca(t) = \{\alpha_{1} (t), \alpha_{2} (t), \ldots, \alpha_{P} (t)\}}$ and it is convenient to write the wavefunction in the form
\begin{equation}
    \Phi(\vecr, \vecat) = \exp \left( \sum_K \locOp_{K}(\vecr) \alpha_{K}(t) \right) ,
    \label{eq:phitVMC}
\end{equation}
where every variational parameter $\alpha_{K}(t)$ is coupled to a local operator
${\locOp_{K}(\vecr)}$~\cite{carleo2012sr}.
For these local operators we use third order B-splines~\cite{boor2001}, which are piecewise polynomial functions, restricted locally to intervals $Y_{mp}$ centered at the points of a uniform grid. We use two sets of intervals, the first ($m=1$) on a uniform grid in $[0,L]$ for the one-body function $u_1$, and the second set ($m=2$) on a grid in $[0, L/2]$ for the two-body correlations $u_2$, where $L$ is the size of the simulation box.
For each interval $Y_{mp}$ we denote the corresponding spline by $B_{mp}(x)$ and
define the corresponding sets of operators ${\locOp_{1p}(\vecr) = \sum_i^N B_{1p}(x_i)}$ and 
${\locOp_{2p}(\vecr) = \sum_{i<j}^N B_{2p}(|x_i - x_j|)}$.
Using this form of the local operators in equation~\eqref{eq:phitVMC}, together with the index mapping $K \equiv (m,p)$ we get
\begin{equation}
    \Phi(\vecr, \vecat) = \exp \left( \sum_p^{P_1} \sum_i^N B_{1p}(x_i) \alpha_{1p}(t) \right) \exp \left( \sum_p^{P_2} \sum_{i<j}^N B_{2p}(|x_i - x_j|) \alpha_{2p}(t) \right) \,.
    \label{eq:phitVMCSums}
\end{equation}
By exchanging the summation in the exponentials we can identify the one- and two-particle correlation functions of the Jastrow-Feenberg ansatz~\eqref{eq:phi0new} as $u_1(x_i, t) = \sum_p^{P_1} B_{1p}(x_i) \alpha_{1p}(t)$ and $u_2(x_i - x_j, t) = \sum_p^{P_2} B_{2p}(|x_i - x_j|) \alpha_{2p}(t)$, respectively.

The effect of the contact interaction in the Hamiltonian~\eqref{eq:hamiltonian} has been
directly incorporated in the wavefunction by using an appropriate boundary
condition on $u_2$ for ${x_i=x_j}$, according to~\cite{lieb1963pr}. In particular, we impose a condition on the variational parameters such that the logarithmic derivative of the wavefunction satisfies ${\frac{1}{\Phi} \frac{\pd}{\pd x_i}\Phi = \frac{1}{4} k_L g}$ for any $x_i = x_j$, which originates from the solution of the two-body problem with contact interaction in one dimension.

As shown in~\cite{carleo2012sr}, the equations governing the time evolution of
the variational parameters are
\begin{equation}
    \ii \sum_{K'} S_{KK'} \dot \alpha_{K'} = 
    \braket{\El \locOp_K} - \braket{\El} \braket{\locOp_K} \,,
    \label{eq:eoms}
\end{equation}
with the correlation matrix
${S_{KK'} = \braket{\locOp_K \locOp_{K'}} - \braket{\locOp_K} \braket{\locOp_{K'}}}$ and  
the local energy ${\El=\frac{H\ket{\Phi}}{\ket{\Phi}}}$.
These coupled nonlinear ordinary differential equations can be solved numerically,
where in every time step the expectation values forming the
coefficient matrix $S_{KK'}$ and the right hand side of the equation
system are calculated by Monte Carlo integration.
In all the simulations presented in this work we use 400 ($P_1 = P_2 = 200$)
complex variational parameters $\alpha_K$, which we have checked to be enough to produce converged results.

%###############################################################################
\subsection{Monte Carlo sampling and time propagation}

In order to accomplish a stable time propagation we need to reduce the numerical errors that are built up during the time evolution of the system. To achieve this, we pre-condition and regularize the matrix $S_{KK'}$ before solving 
the Eqs.~\eqref{eq:eoms}.  As a first step we scale the matrix by ${S_{KK'}' = S_{KK'}/\sqrt{S_{KK}S_{K'K'}}}$ 
and as a second step we add a small 
regularizing factor 
$\varepsilon$ to the diagonal entries 
(${S_{KK'}' \to S_{KK'}' + \varepsilon \delta_{KK'}}$) 
in order to prevent instabilities due to eigenvalues that are close to zero in 
$S_{KK'}'$~\cite{becca2017}. 
The same value ${\varepsilon = 10^{-4}}$ was used in all simulations. 
To solve the resulting system of equations 
we use a QR decomposition and a fourth order Runge-Kutta scheme to propagate the differential equations~\eqref{eq:eoms} in time. We found that a stable time propagation requires a reasonably small time step of at least $\delta t = 10^{-4}\,t_0$, which we used throughout this work. The Monte Carlo estimates for 
$S_{KK'}, 
\braket{\El \locOp_K}, \braket{\El}$ and $\braket{\locOp_K}$ are 
obtained 
using the Metropolis-Hastings algorithm, and a total of ${N_{\text{MC}} = 12500}$ uncorrelated samples are used in every time step of the numerical propagation of 
Eq.~\eqref{eq:eoms}. 
The density observable $\braket{\rho(r, t)}$, which is the main quantity of interest in our simulations, is calculated at every hundredth simulation time step, i.e. at multiples of a time step ${\delta t_\rho = 0.01\,t_0}$. In order to get the density estimate with high accuracy we use ${N_{\text{MC}, \rho} = 2.5 \cdot 10^6}$ uncorrelated samples.

%###############################################################################
%###############################################################################
\begin{figure}
	\centering
    \includegraphics{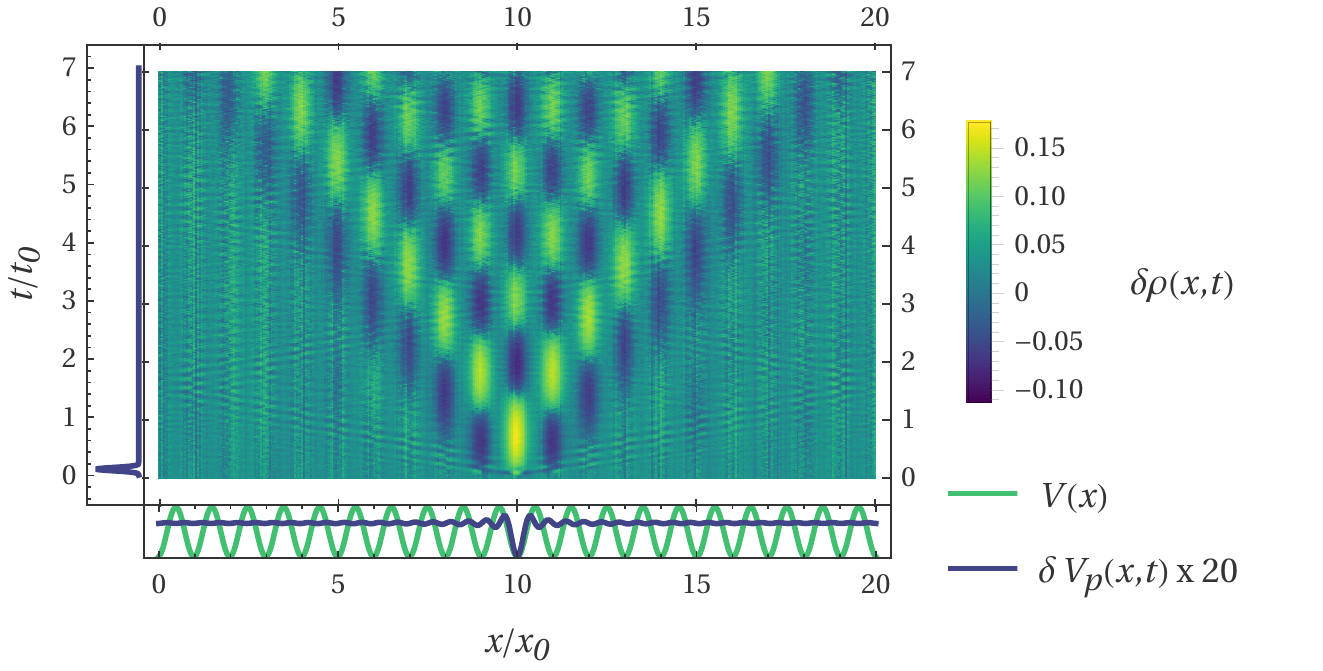}
    \caption{\label{fig:simulationSketch} 
    Spatial density fluctuation $\delta \rho(x,t) = \rho(x,t) - \rho(x,0)$ evolving in time,
    from which we obtain $S(k,\omega)$.
    Initially the system is in the ground state (obtained via i-tVMC) of the
    optical potential $V(x)$ (green line). At time $t=0$, a weak
    perturbation pulse $\delta V_p(x,t)$ with a Gaussian time profile (blue
    line along vertical axis) and a superposition of various momentum modes 
    (blue line along horizontal axis), given by equation~\eqref{eq:pulse}, is turned on.
    The main color map shows the propagation in time (vertical) and space (horizontal axis) of the density
    fluctuation $\delta \rho$ induced by the perturbation.
    The amplitude of the pulse $\delta V_p$ is magnified by a factor of 20.
    The system parameters ${V_0=7\,E_r}$, ${g=0.41\,E_r/k_L}$ (corresponding to ${U/J = 6}$) and the pulse parameters ${V_e=0.0125\,E_r}$, ${t_e=0.1\,t_0}$ and ${\tau=0.04\,t_0}$ are used in this simulation.
    }
\end{figure}

\section{Simulations}

For the calculations we proceed as follows: we first perform tVMC simulations in
imaginary time (i-tVMC) with $\delta V_p=0$ to obtain the variational ground state of the
Hamiltonian in Eq.~\eqref{eq:hamiltonian}.
The result is then used as the initial wavefunction for the real time
simulation, where we turn on the perturbing potential $\delta V_p$ at $t=0$ and
monitor the density fluctuations $\delta \rho(x, t)$ (see
\figref{fig:simulationSketch}). If the perturbation is weak, we use linear response
theory~\cite{pines1966} to estimate the dynamic structure factor 
\begin{equation}
    S(k, \omega) = -\frac{1}{\pi} \operatorname{Im} \left[ \frac{\delta\tilde\rho(k, \omega)}{\delta \tilde V_p(k, \omega)} \right] \,,
    \label{eq:skw}
\end{equation}
where ${\delta\tilde\rho(k, \omega)}$ and ${\delta \tilde V_p(k, \omega)}$ 
are the space and time Fourier transforms of the density fluctation and the perturbing
potential, respectively.

For comparison with exact ground state results, we also performed i-tVMC calculations 
in the absence of the optical lattice (${V_0 = 0}$), leading to 
the Lieb Liniger model~\cite{lieb1963pr}. The resulting ground state energy compares well to the energy obtained in Bethe ansatz calculations (see Ref.~\cite[Eq.~(10)]{lang2017sp}), with a relative error of less than 0.4\% for the range of interaction strengths~$g$ used in this work.

%###############################################################################
\subsection{Linear response}
\label{ssec:linrep}

We calculate the dynamic structure factor $S(k,\omega)$ from Eq.~\eqref{eq:skw} for several values of
the coupling strength~$g$ and the optical lattice amplitude~$V_0$.
We use $N=20$ particles with a density $n=1/x_0$ in a simulation box of size
$L=x_0 N$ with periodic boundary conditions, corresponding to unit filling.
To excite the system we apply a multi-mode pulse with a Gaussian time profile 
\begin{equation}
    \label{eq:pulse}
    \delta V_p(x, t) = V_e \, e^{ - \left( t-t_e \right)^2/\tau^2} \
    \sum_{j}^{j_\text{max}} \sin^2 \left( k_j x \right) ,
\end{equation}
where the spatial part is a superposition of up to ${j_\text{max}=40}$ modes
with wave numbers given by ${k_j = 2 \pi j / L}$.
In particular, we choose ${V_e=0.0125\,E_r}$, ${t_e=0.1\,t_0}$ and
${\tau=0.04\,t_0}$. This pulse imparts an energy less than
$0.25\%$ of the ground state energy to the system, which shows that the perturbation is weak enough for linear
response theory to apply. To check this further we doubled~$V_e$ and indeed got the same $S(k,\omega)$.
In the linear regime we can get the full excitation spectrum in a single tVMC
simulation since modes are excited simultaneously, but independently of each other.
The short pulse length $\tau$ also ensures that it excites a broad range ${2\pi/\tau}$ of energies.
In any case, the pulse in Eq.~\eqref{eq:pulse} can be easily tailored, to
excite only selected modes if required.

\begin{figure}
	\centering
    \includegraphics{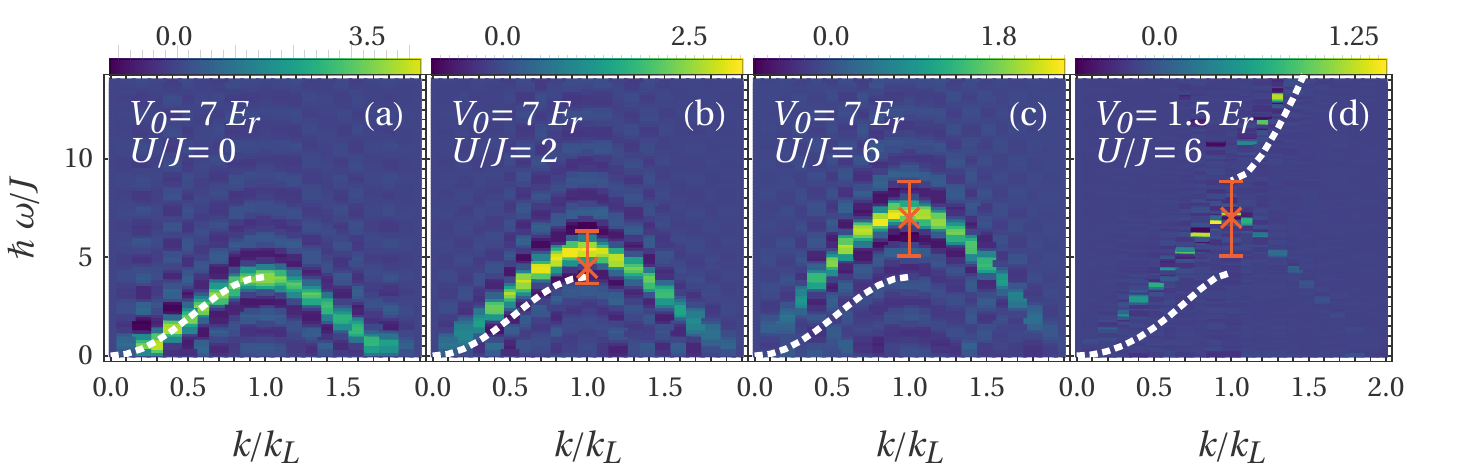}
    \caption{\label{fig:skwLinearResponse}
    Dynamic structure factor $S(k,\omega)$ from tVMC simulations for bosons
    in an optical lattice of amplitude ${V_0=7\,E_r}$ and equivalent BHM parameter
    ${U/J = 0;2;6}$ (panels (a)--(c)), and $V_0=1.5\,E_r$ with $U/J=6$
    (panel~(d)).
    The dashed white lines show the dispersion of non-interacting particles
    ($U=0$) for the given value of~$V_0$, obtained from band structure
    calculations.
    The red bar indicates the spread of the multipeak feature in $S(k,\omega)$ of
    Ref.~\cite{roux2013njop}.
    }
\end{figure}

We present in \figref{fig:skwLinearResponse} the dynamic structure factor
$S(k,\omega)$, 
in units of $k_L$ and $J/\hbar$ for~$k$ and~$\omega$, respectively.
Panels~\mbox{(a)--(c)} show $S(k,\omega)$ for a deep optical potential
$V_0=7\,E_r$ and interaction strengths $g=0;0.14;0.41\,E_r/k_L$,
corresponding to the ratios $U/J=0;2;6$ of the BHM, respectively. 
Panel~(d) shows $S(k,\omega)$ for a shallow lattice with $V_0=1.5\,E_r$ and 
$g=2.8\,E_r/k_L$, corresponding to the equivalent BHM ratio $U/J=6$.
White dashed lines denote the Bloch dispersion of non-interacting particles.
The tVMC result for $U/J=0$ in panel~(a) demonstrates that the peaks in
$S(k,\omega)$ reproduce the exact non-interacting Bloch dispersion perfectly.
The broadening of the tVMC dispersion, as well as the ringing
oscillations, are artifacts resulting from the Fourier transform over 
a finite simulation time of length $T=10\,t_0$.
When we increased $T$ and thus the computational cost, the artificial oscillation frequency increased and the amplitude decreased.
As $U/J$ is increased, the excitation energies increase also, and the dispersion
becomes linear for small~$k$.
The positions of the peaks in $S(k,\omega)$ as function of $\omega$ are in good agreement with results
of~\cite{roux2013njop} obtained by exact diagonalization of the BHM.
The details of our $S(k,\omega)$, however, differs from the results
in~\cite{roux2013njop}, where multiple close peaks were obtained for $N=16$.
In panels (b)~to~(d), the spread of these peaks is indicated by a red bar, with the
central main peak of \cite{roux2013njop} indicated by a cross.
The main difference of our system compared to~\cite{roux2013njop} is that we use continuous coordinates instead of using the Hubbard approximation leading to the discrete lattice of the single-band BHM. Furthermore we simulate a slightly higher number of particles and use a variational description of the wavefunction.

For the shallow optical lattice case shown in panel~(d), corresponding to
${V_0=1.5\,E_r}$, the band gap is comparable with the band width.
In such a case the single-band BHM 
does not apply.
The dispersion is linear over a wider range of $k$ values than in the deep
lattice, while the maximum of the first band hardly changes.
In such a shallow lattice, we also observe an energy increase of the second band
with respect to the Bloch dispersion.
Notice that the seemingly smaller broadening of the curve in panel~(d) is due to
the fact that all energies are expressed in units of $J$, which is $J=0.04\,E_r$
for $V_0=7\,E_r$ and $J=0.16\,E_r$ for $V_0=1.5\,E_r$.

%###############################################################################
\subsection{Nonlinear response}

\begin{figure}
	\centering
    \includegraphics{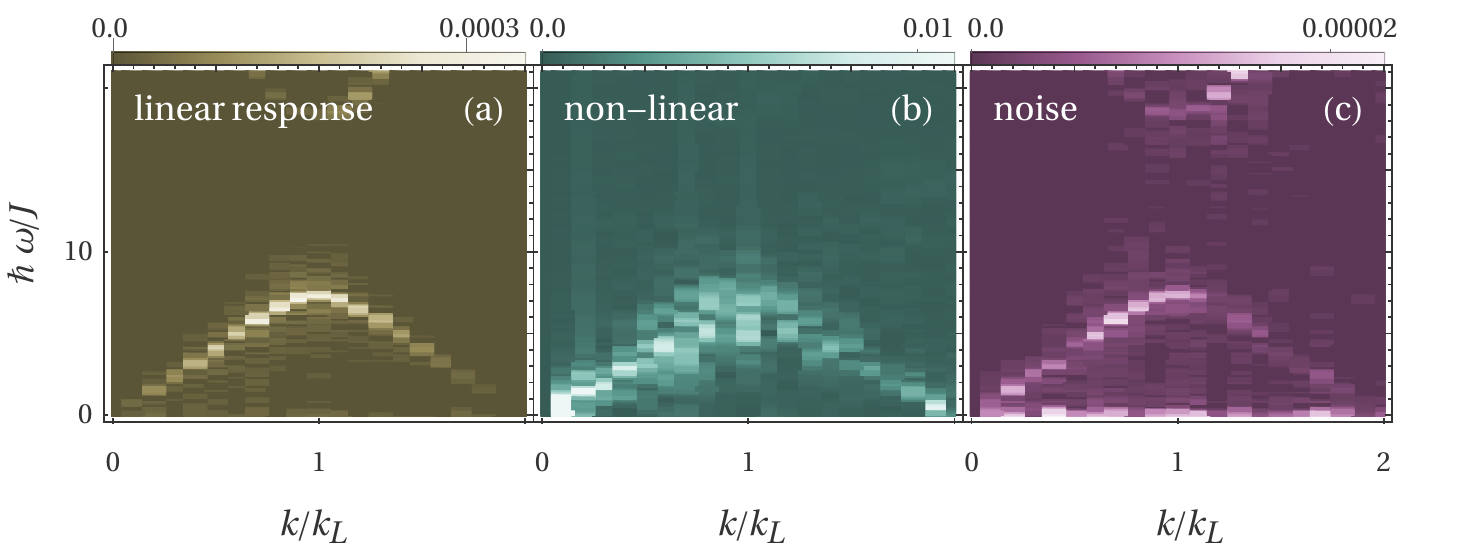}
    \caption{\label{fig:pulseAndNoise} 
    Square root of the power spectrum $|\delta \tilde \rho(k, \omega)|$ of the density
    fluctuations obtained with three different tVMC simulation variants. 
    Panel (a) shows the response to a weak multi-mode pulse of the form
    given in Eq.~\eqref{eq:pulse}.
    In panel (b), the system is excited with a single mode at
    $k=0.1\,k_L$ with an amplitude as large as the lattice potential
    ($V_e=V_0$).
    In panel (c), no perturbation is applied and the density fluctuations in the
    time propagation are solely due to the stochastic noise in the Monte Carlo
    simulation. In all three simulations
    the lattice amplitude is $V_0=3\,E_r$ and the BHM parameter is $U/J=6$.
    }
\end{figure}
The tVMC method is not restricted to weak perturbations, and thus one
can use it to explore the response of the system outside the linear regime.
In order to demonstrate this, we again perturb the same Bose system at unit
filling in the optical lattice with ${V_0=3\,E_r}$ and ${U/J=6}$, but this time
with a strong pulse. 
Instead of exciting all wave numbers simultaneously with the weak pulse in
Eq.~\eqref{eq:pulse}, we excite only the lowest mode compatible with the
periodic boundary condition, with wave number
${k_1=2\pi/L=0.1\,k_L}$, but with a pulse strength equal to the amplitude of the
lattice potential, ${V_e=V_0}$.
We use a pulse length $\tau$ five time longer than in the linear response
simulations previously described, and also set $t_e=0.5\,t_0$ to move the peak
of the pulse to larger times for a smooth switch-on of the perturbation.
Overall, the integrated pulse strength is~30 times
stronger than that of the weak multi-mode pulse to compare with.
Outside the linear regime, $S(k,\omega)$ no longer describes the full
response of the system to the perturbation, and furthermore, for $k\ne k_1$
we have $\delta\tilde V_p(k,\omega)=0$, and thus $S(k,\omega)$ cannot even be calculated.
Therefore we show the square root of the power spectrum, ${|\delta\tilde \rho(k, \omega)|}$. 

The panel (b) of \figref{fig:pulseAndNoise} shows ${|\delta\tilde \rho(k, \omega)|}$
after the strong pulse with wave number $k_1$. 
As it could be expected, a very pronounced peak in the non-linear response
appears at $k_1$. However, the strong pulse excites a wide range of multiples
of $k_1$ via higher harmonic generation.
For comparison, in panel (a) of \figref{fig:pulseAndNoise} we
show ${|\delta\tilde \rho(k, \omega)|}$ for the weak multi-mode pulse
$\delta V_p(x, t)$ of Eq.~\eqref{eq:pulse}.
Note that in the linear response regime ${|\delta\tilde \rho(k, \omega)|}$
conveys the same information as $S(k,\omega)$,
see the previous~\figref{fig:skwLinearResponse}.
Compared to the linear response to the weak multi-mode pulse, the non-linear
response exhibits a much broader excitation band, but it essentially follows the
dispersion relation obtained from linear response; the broadening is expected
for higher harmonic generation in a system with a non-linear dispersion.
Panel (b) of \figref{fig:pulseAndNoise}
demonstrates that
a sufficiently strong long wavelength perturbation yields the full
excitation spectrum, albeit with significant broadening.

%###############################################################################
\subsection{Excitations from noise}

An even more remarkable feature of the tVMC method is that the full
excitation spectrum can also be obtained in the opposite limit, i.e.\
applying no perturbation at all. We can simply propagate the variational ground state
in real time. The stochastic noise in $S_{KK'}$ and the right hand
side of Eq.~\eqref{eq:eoms}
produces fluctuations around the exact time evolution which we can use
to calculate the excitation spectrum of all modes. 
Similarly to the nonlinear case,
$S(k,\omega)$ is not accessible because $\delta \tilde V_p(k, \omega)=0$,
this time for all $k$.
We show ${|\delta\tilde \rho(k, \omega)|}$, generated entirely by the stochastic noise,
in panel (c) of~\figref{fig:pulseAndNoise}.
The peak locations giving the excitation energies are essentially identical
to the linear response results shown in panel (a).
In this way, the Monte Carlo noise can be effectively used to explore 
the excitation spectrum of the system, although as seen from the color
scales in~\figref{fig:pulseAndNoise},
the noise generated power spectrum is much weaker.

As expected, the noise is reduced when we increase the sample size per time step,
but the signal-to-noise ratio of the density fluctuation power spectrum remains
unchanged. If, on the other hand, we improve the variational ansatz $\Phi$, the parameter optimization
with i-tVMC leads to a variational ground state closer to the exact ground state. When we increased
the number of parameters~$\alpha_K$, the signal-to-noise ratio in ${|\delta\tilde \rho(k, \omega)|}$
dropped, because improving the variational wave function reduces the variance of the local energy $\El$.
The sampling noise in the quantities on the right hand side of Eq.~\eqref{eq:eoms} falls, while the sampling noise in
$S_{KK'}$ on the left hand side 
is barely affected.
In this way, there 
are less noise-induced perturbations to the ground state evolution
of the parameters $\alpha_K$ when we solve Eq.~\eqref{eq:eoms}. In the limit that the optimized
ansatz~$\Phi$ is the \emph{exact} ground state, $\El$ is the exact ground state energy, with zero
variance, while the correlation matrix $S_{KK'}$ is still non-zero and invertible, which leads to $\dot\alpha_K=0$, thus there is no noise-induced time evolution. The only noise left
in the power spectrum of the density fluctuation is the sampling noise which carries no
information on the dynamics because it is uncorrelated between time steps.
But apart from a few selected problems, the exact many-body wave function is not known, and
in general there will always be some noise-induced time evolution of $\alpha_K$ about
their optimized values.

Our result in panel (c) of~\figref{fig:pulseAndNoise} suggests a new simulation strategy where the unperturbed
ground state is propagated in time, rather than exciting specific modes 
with suitable temporally and spatially shaped weak or strong external pulses.
With this new type of simulation we can for example determine the
excitation spectrum in a large range of $\omega$- and $k$-values,
which is useful when analyzing a new system with little
knowledge about the relevant range of energies and momenta
to explore. It has the added benefit of not having to choose any specific
form for the perturbation potential.
We stress, however, that ${|\delta\tilde \rho(k, \omega)|}$ is not proportional to
$S(k,\omega)$. We can obtain the excitation energies from the peaks of either of
them, but the stochastic noise is not white and thus has different strength for different
energies and momenta. For example, in the present case, the peaks in panel~(c)
of~\figref{fig:pulseAndNoise} are clearly smaller for $k>k_L$ than for $k<k_L$, while the linear
response result in the left panel looks more symmetric about $k_L$.
In order to obtain the spectral weights of the dynamic structure function, we
have to use linear response theory as demonstrated in section~\ref{ssec:linrep}.

%###############################################################################
%###############################################################################
\section{Conclusion}

In summary, we have explored the possibility of using time-dependent variational
Monte Carlo (tVMC) to obtain the dynamic structure
factor $S(k,\omega)$, or more generally the excitation spectrum, of
many-body quantum systems under the action of a pulsed perturbation.
Specifically, we have analyzed the linear and nonlinear dynamics of a
one-dimensional system of bosons in an optical lattice described by a
continuous Hamiltonian.
In both deep and shallow lattices, we explore several interaction strengths
corresponding to the same ratio $U/J$ of the Hubbard interaction and hopping
parameters, to assess the universality of the dependence of the excitation
spectrum on it.
For shallow lattices and as expected, we observe a deviation from
the single-band Bose-Hubbard result, with the dispersion being linear over a
wider range of momenta.
However, for the lowest band, the excitation energy at the edge of the Brillouin
zone is remarkably universal.

Besides the weak perturbation regime where linear response theory applies, we
have also explored the dynamics after a very strong perturbation, and the
dynamics with no perturbation at all.
In the latter case we simply propagate the optimized ground state in real time
to obtain the excitation spectrum from the fluctuations due to the
stochastic noise intrinsic to every Monte Carlo method.
This can be useful when studying complex systems where the nature of the 
excitations is not known, and the right choice of the perturbation operators is not so obvious.
In order to explore the non-linear regime, we apply pulses coupling to a single
mode, but with peak strengths of the order of the optical lattice depth itself. 
These strong pulses excite the full range of wave numbers via higher harmonic
generation.
This could be relevant for Bragg spectroscopy of the excitation spectrum, since only
one or few momentum transfers need to be chosen to obtain an
approximate $S(k,\omega)$ for a wide range of momenta.

\section*{Acknowledgments}
M.\,G. and R.\,E.\,Z thank G.~Carleo and \mbox{M.~Holzmann} for fruitful discussions
and acknowledge computational resources of the Scientific Computing
Administration at Johannes Kepler University.

\paragraph{Funding information}
F.\,M. acknowledges financial support by grant \mbox{PID2020-113565GB-C21} funded
by \mbox{MCIN/AEI/10.13039/501100011033}, and from Secretaria d'Universitats i
Recerca del Departament d'Empresa i Coneixement de la Generalitat de Catalunya,
co-funded by the European Union Regional Development Fund within the 
ERDF Operational Program of Catalunya (project QuantumCat, 
ref.~\mbox{001-P-001644}).

\bibliography{literature}

\end{document}